\documentclass[twocolumn]{aastex631}

\usepackage{amsmath}
\usepackage{color}

\newcommand{\kms}{km\,s$^{-1}$}
\newcommand{\re}[1]{#1}
\newcommand{\mg}[1]{#1}
\newcommand{\bl}[1]{#1}
\newcommand{\bn}[1]{#1}
\newcommand{\rv}[1]{#1}
\newcommand{\rvb}[1]{#1}

\shorttitle{H$_3^+$ pure rotational lines in HH 168}
\shortauthors{Neufeld et al.}

\hypersetup{hidelinks}

\begin{document}
\nocite{Neufeld1989a}

\title{First detection of the forbidden rotational spectrum of H$_3^+$:\\ JWST/MIRI observations of the shocked gas in HH 168}

\author[0000-0001-8341-1646]{David A. Neufeld}
\affiliation{William H. Miller III Department of Physics and Astronomy, Johns Hopkins University, 3400 North Charles Street, Baltimore, MD 21218, USA}
\email{neufeld@jhu.edu}

\author[0000-0003-1665-5709]{Joel D. Green}
\affiliation{Space Telescope Science Institute, 3700 San Martin Drive, Baltimore, MD 21218, USA}

\author[0000-0001-8302-0530]{Dan M. Watson}
\affiliation{Department of Physics and Astronomy, University of Rochester, Rochester, NY 14627, USA}

\author[0000-0003-2027-5020]{Paule Sonnentrucker}
\affiliation{European Space Agency (ESA) Office, Space Telescope Science Institute, 3700 San Martin Drive, Baltimore, MD 21218, USA}

\begin{abstract}
The H$_3^+$ molecular ion is the cornerstone of interstellar ion-molecule chemistry. Its vibrational transitions have long been observed, in
absorption along many interstellar sight lines and in emission from the upper atmospheres of the giant planets, but its pure
rotational spectrum, which consists only of ``forbidden'' transitions made possible by centrifugal distortion, had not been
observed in the laboratory or in any astronomical source.
\bl{With JWST/MIRI} we have detected this spectrum
toward the Herbig-Haro object HH 168 in Cepheus A West. \bn{Twenty-six} lines are detected above 5\,$\sigma$ in one or both of two fields,
of which the $(J,K) = (5,0) \to (4,3)$ line at $16.325\,\mu$m is the strongest. The H$_3^+$ lines share the velocity
and spatial distribution of the hot shocked H$_2$, not of the ionized gas. A rotation diagram yields
a rotational temperature of $\approx$\,\bl{730}\,K, but the level populations are not thermal and are best explained by
subthermal excitation in gas at $\gtrsim 2000$\,K. We compare the observations with models of shocks
propagating into mildly irradiated molecular gas. \re{Constrained by the H$_2$ rotational lines, a dissociative J-type shock predicts
H$_3^+$ emission within a factor $\sim 1.5$ of that observed, while a nondissociative C-type shock, which fits the H$_2$ lines
better, underpredicts it by a factor \bl{5--6}}, although the excitation of H$_3^+$ by energetic collisions is poorly known. \bl{Archival MIRI data show the H$_3^+$ forbidden rotational spectrum in the HH 211 bow shock. The H$_3^+$ lines} provide a new probe of the ionization in shocked molecular gas.
\end{abstract}


\section{Introduction} \label{sec:intro}

The H$_3^+$ ion is formed when cosmic rays ionize H$_2$ and the resulting H$_2^+$ reacts with H$_2$; it then donates a
proton to other species, initiating the ion-molecule chemistry of molecular clouds. Because H$_3^+$ is an
equilateral triangle, it has no permanent dipole moment and no allowed pure rotational spectrum. Its near-infrared
vibrational spectrum has been observed in emission from the ionosphere of Jupiter \citep{Drossart1989} and in absorption
in interstellar space \citep{Geballe1996}, where it has provided a key means of determining the cosmic-ray ionization rate in
\re{diffuse molecular} clouds \citep[e.g.,][]{McCall2003,Oka2006,Indriolo2012}.

\rv{Molecules that lack a permanent dipole moment can nevertheless have weak ``forbidden'' rotational spectra, through a dipole moment induced by centrifugal distortion \citep{Watson1971}.}
\bl{As calculated by \citet{Pan1986},}\footnote{\rv{An earlier calculation by \citet{Aliev1984} gave Einstein $A$ coefficients about seven times smaller than those of \citet{Pan1986}; the latter agree with the modern ab initio values adopted here \citep{Mizus2017} to within a factor of 1.6.}}
centrifugal distortion of the rotating H$_3^+$ ion induces a small dipole moment and gives rise to weak ``forbidden''
transitions with $\Delta k = \pm 3$ within the ground vibrational state. These lie in the mid-infrared,
with Einstein $A$ coefficients \re{of up to $\sim 0.5\,\rm s^{-1}$}. Because H$_3^+$ is light and rotates rapidly, the distortion
dipole is unusually large: \re{the $A$ coefficients of the lines detected here ($7 \times 10^{-4}$--$0.5\,\rm s^{-1}$)} exceed those of the H$_2$ quadrupole lines
($A \sim 10^{-10}$--$10^{-7}\,\rm s^{-1}$) at similar wavelengths by five to eight orders of magnitude. \re{Although Pan \& Oka noted that the forbidden rotational spectrum provided
``an alternative possibility of detecting interstellar H$_3^+$,'' it remained unobserved over the following four decades,}
either in the laboratory or in any astronomical source \citep{Miller2020}. A search for the
rotational lines in Jupiter's auroral emission was unsuccessful \citep{Trafton2009}, and most recently, \citet{Goicoechea2025} predicted that its strongest line, $(J,K) = (5,0) \to (4,3)$
at $16.325\,\mu$m, might be detectable with JWST, and placed an upper limit on it toward the Orion protoplanetary disk
d203-506. Here we report the detection of this spectrum, through many lines, toward the Herbig-Haro object HH 168.

\section{Observations and data reduction} \label{sec:obs}

HH 168 is a bright Herbig-Haro object in Cepheus A West, at a distance of $\approx 700$\,pc \citep{Moscadelli2009}.
At high angular resolution it resolves into a shocked bubble bounded by a series of bow shocks seen in hydrogen and ionic
emission lines \citep{Hartigan2000}. The proper motions of its high-excitation knots \bl{reach} $\sim 200$\,\kms\bl{, and their optical emission lines are several hundred \kms\ wide, whereas} the radial velocities of the
shocked H$_2$ lie within $\sim 30$\,\kms\ of the cloud velocity \citep[][and references therein]{Hartigan2000}\bl{: the H$_2$ traces slower shocks along the periphery of the fast flow. HH 168 lies}
far from any confounding stellar emission; the detection of strong [Ne\,{\sc iii}]\,15.56\,$\mu$m emission in our
spectra indicates that some of its shocks are considerably faster than those that produce the H$_2$ emission.

It was observed with the Medium Resolution Spectrometer \bn{\citep[MRS;][]{Wells2015,Argyriou2023}} of the Mid-Infrared Instrument (MIRI) as part of JWST GO
program 7084 (``Measuring the Thickness of a J-Shock in Cep A West''; PI: J.~D.~Green), covering
$4.9$--$27.9\,\mu$m at a resolving power of $\sim$1300--3700. Two fields along the shock, C and G, were observed on 2026
August 1 (observations 1 and 2). Each was mapped at five pointings with the 4-point dither pattern in all three grating
settings, with the FASTR1 readout pattern, 57 groups and 2 integrations per exposure (319\,s), for a total of 6383\,s per
setting; a dedicated background field was observed in the same settings (observation 3).
The \re{calibrated exposures retrieved from MAST (pipeline version 2.0.1) were processed through stage 3} with version 3.0.0 of the JWST pipeline (Calibration Reference Data System context {\tt jwst\_1535.pmap}), the background being
subtracted from the calibrated exposures (master background) before the spectral cubes were built. The data are available at the Mikulski Archive for Space Telescopes (MAST): \dataset[doi:10.17909/6wh2-6424]{https://doi.org/10.17909/6wh2-6424}.

We analyze two fields, G and C (named after the knots G and C of \citealt{Hartigan1985}; see \citealt{Hartigan2000}), in each of which we extract a spectrum over a contiguous region of $\approx
3\,\rm arcsec^2$ chosen to maximize the H$_2$ S(7) intensity, with all lines measured at a common angular resolution. For each line we
made a map from Gaussian fits, spaxel by spaxel, \bn{of fixed center and instrumental width plus a quadratic baseline, over a window of $\pm 5$ FWHM. The width was measured in each MIRI sub-band from the brightest unblended lines, and the bright H$_2$ $v = 0$ lines were fitted freely. The adopted widths correspond to resolving powers of 3300--3500 in channels 1 and 2, 3000--3200 in channel 3 and 2100--2300 in channel 4, higher in channel 4 than the in-flight values of \citet{Argyriou2023}, as also found by \citet{Pontoppidan2024b}. Cataloged lines of other species, and other H$_3^+$ lines, within the fit window were included as additional fixed-center components. Each map was convolved with a Gaussian of FWHM $(\theta_{23}^2 - \theta_\lambda^2)^{1/2}$, where $\theta_\lambda$ is the width of the MRS
point-spread function at the line wavelength and $\theta_{23} = 0.87''$ its value at $23\,\mu$m, and regridded to a common
$0.35''$ grid. The line flux is the sum of the convolved map over the 3.06\,arcsec$^2$ aperture, which was chosen on the convolved S(7) map. Every line shortward of $23\,\mu$m, including all those in Table~\ref{tab:lines}, is therefore measured with the same effective beam, and no aperture correction is needed. The few lines at longer wavelengths are measured at their native resolution ($\lesssim 1''$).} Uncertainties follow from the
residuals of the same fit to the aperture spectrum of the convolved cube.

\section{Results} \label{sec:results}

\subsection{Extinction correction} \label{sec:ext}

We determined the extinction from the H$_2$ rotational lines themselves\mg{, as in, e.g., \citet{Neufeld2024} for the protostar HOPS 370}: fitting the H$_2$ S(1)--S(8)
fluxes with a two-temperature model in which the ortho-to-para ratio of each component is free, together with the KP5
extinction curve \citep{Chapman2009,Pontoppidan2024}, we obtain $A_V = $\,3.0 (G) and 4.1\,mag (C), with ortho-to-para
ratios of 1.8 and 1.7 for the warm gas. \mg{The H$_2$ rotation diagram is curved, as expected for shocked gas that spans a range of temperatures; the two-temperature model is simply a convenient description of this curvature, used here only to determine $A_V$, and does not necessarily imply two physically distinct gas components.} Exactly the same extinction and extinction curve were then used to correct the
H$_3^+$ line fluxes. Forcing an ortho-to-para ratio of 3 in both temperature components raises $A_V$ to 3.7 and 5.6\,mag but fits the H$_2$ lines much worse; even this
larger extinction would change the corrected H$_3^+$ intensities by less than 0.1\,dex.

\subsection{Detection} \label{sec:detect}

\begin{figure*}[t]
\plotone{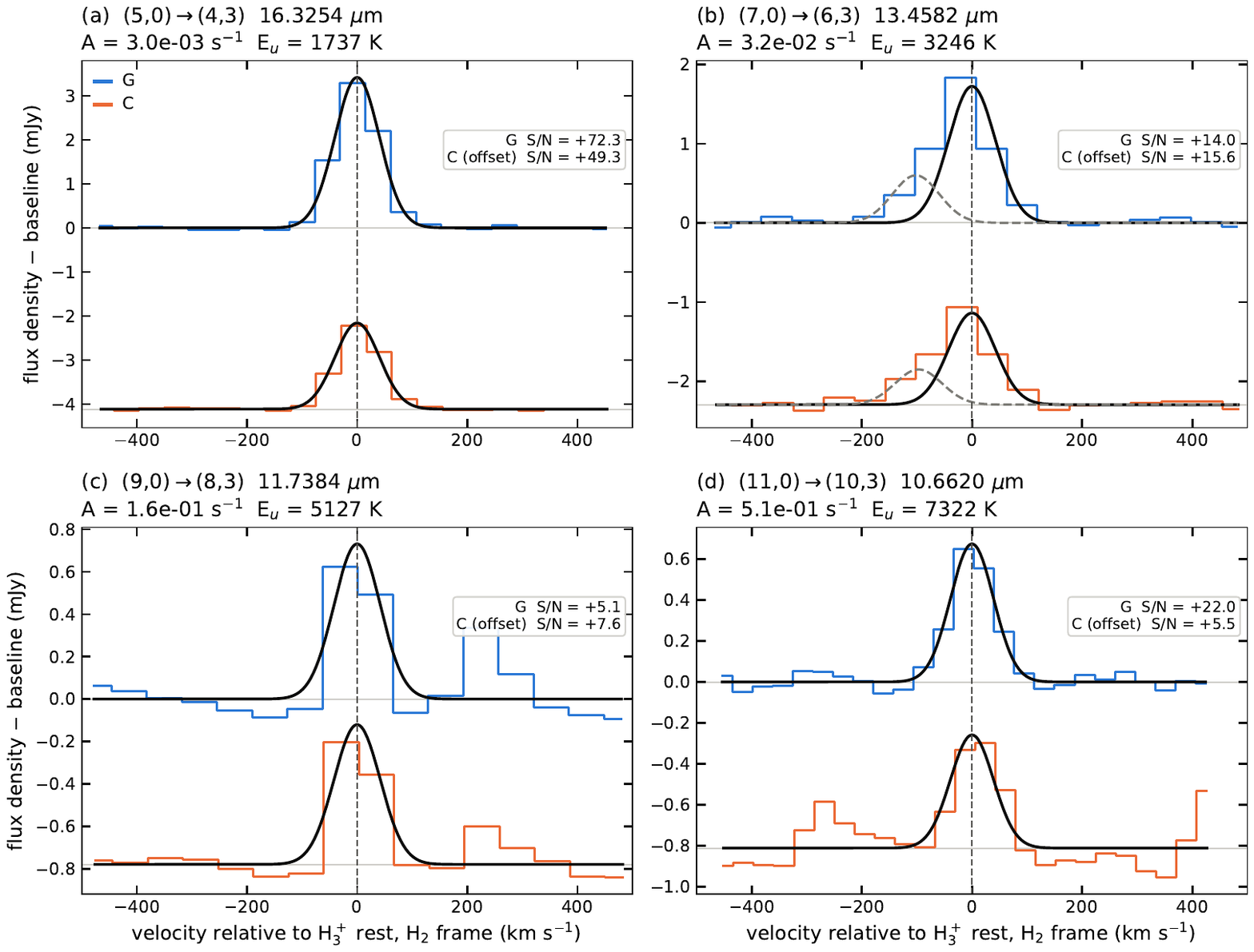}
\caption{The $(J,0) \to (J-1,3)$ series of H$_3^+$ forbidden lines for $J = 5$, 7, 9 and 11 (panels a--d) in fields G (blue, top)
and C (orange, offset downwards), plotted against velocity relative to the H$_3^+$ rest wavelength in the frame of the H$_2$ \bn{S(7)} emission,
after subtraction of a quadratic baseline\bn{, for the aperture spectra of the convolved cubes from which the fluxes of Table~\ref{tab:lines} are measured}. Black curves: Gaussians of fixed center and instrumental width\bn{; dashed: a neighboring line fitted simultaneously}.
The (5,0)\,$\to$\,(4,3) line at $16.325\,\mu$m was predicted to be the strongest by \citet{Goicoechea2025}.
\label{fig:lines}}
\end{figure*}

We measured the fluxes of all 96 $v = 0$ forbidden lines of H$_3^+$ with $A \ge 10^{-4}\,\rm s^{-1}$ and upper-level
energies below $6000\,\rm cm^{-1}$ ($E_u/k < 8630$\,K\rv{; throughout, level energies are measured from the lowest level, (1,1), whereas the MiZATeP states file lists them relative to the nonexistent (0,0) level, for which $E(1,1) = 64.12\,\rm cm^{-1}$}) that fall \re{at 7.6--26.3\,$\mu$m}, using the ExoMol MiZATeP line list \citep{Mizus2017}
with empirical energy levels from the MARVEL (measured active rotational-vibrational energy levels) analysis \citep{Furtenbacher2013}\rvb{, as implemented in the updated ExoMol line list of \citet{Bowesman2023}}\rv{.} \rvb{\citet{Markus2019} determined the ground-state energies to high accuracy from laboratory measurements; their predicted frequencies for the forbidden transitions agree with those adopted here to within 1 part in $10^6$ at mid-infrared wavelengths}. \bn{Twenty-six} lines that do not coincide, within the instrumental resolution, with known lines of other species are detected at more than $5\,\sigma$ \mg{in at least one of the two fields}
(Table~\ref{tab:lines}): \bn{25} lines, from \bn{22} upper states with $E_u/k = $\,1700--7300\,K, in G and \bn{19} in C\bl{, \bn{18} of them in both}.
The lowest-lying upper level accessible to MIRI, $(4,1)$ at $E_u/k = 1107$\,K, is
tentatively detected through the $(4,1) \to (3,2)$ line at $24.657\,\mu$m (\bl{$3.7\,\sigma$} in G, \bl{$1.3\,\sigma$} in C); all other lines
from levels with $J \le 4$ lie at wavelengths beyond $27.9\,\mu$m. Figure~\ref{fig:lines} shows the $(J,0) \to (J-1,3)$ series for $J = 5$, 7, 9
and 11. The $(5,0) \to (4,3)$ line at $16.325\,\mu$m, the strongest predicted by \citet{Goicoechea2025}, is detected at
\bl{$72\,\sigma$} in G and \bl{$49\,\sigma$} in C, with a flux of \bl{$2.1 \times 10^{-16}$}$\,\rm erg\,cm^{-2}\,s^{-1}$ in G.

HH 168 proves to be a favorable source for detecting these lines. Lying far from the embedded protostars of Cep A, it has a line-poor
mid-infrared spectrum, dominated by H$_2$, atomic fine-structure and H\,{\sc i} lines, with few lines of other molecules;
the weak H$_3^+$ lines therefore suffer little interference from other species.

Several tests support the identification. Fitting all lines with a common velocity offset, the H$_3^+$ lines are at the
velocity of the H$_2$ emission (\mg{\bl{$+1.0$} and \bl{$-4.5$}\,\kms\ relative to the mean of H$_2$ S(1)--S(4) in G and C; the H$_2$ lines themselves differ by up to 10\,\kms\ between MIRI sub-bands, a measure of the uncertainty in the wavelength calibration}). Most lines that share
an upper level give consistent populations. 

\begin{figure*}[t]
\plotone{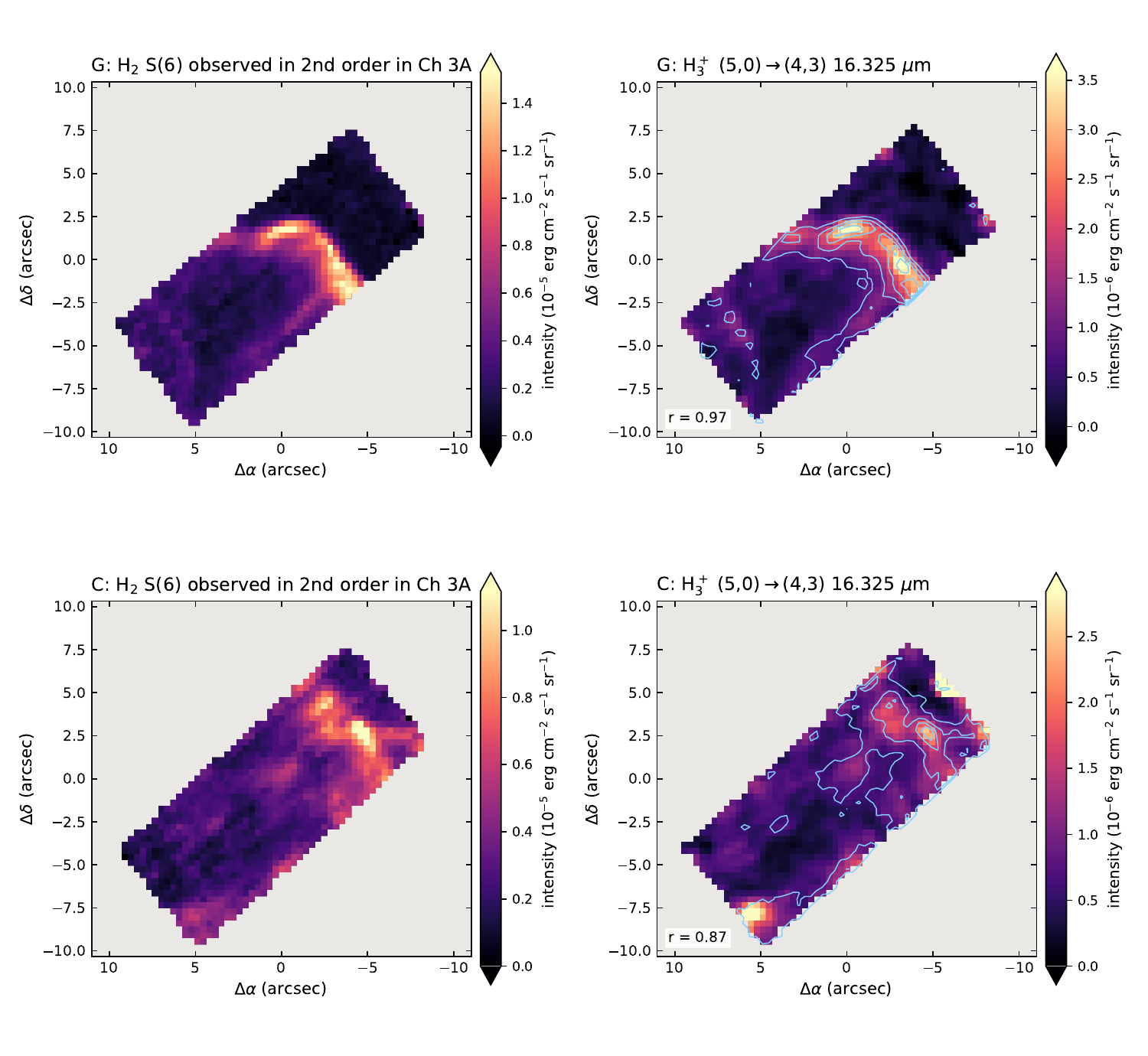}
\caption{Maps of H$_2$ S(6) observed in second order in MIRI channel 3A (left; this second-order image carries 1.5\% of the first-order S(6) flux,
\citealt{Neufeld2026}; shown without further convolution, \re{its resolution, measured on these maps as $\approx 0.86''$ (larger than the nominal $0.51''$ adopted by \citealt{Neufeld2026}), matches that of the H$_3^+$ map}), and of the H$_3^+$ (5,0)\,$\to$\,(4,3) line at $16.325\,\mu$m (right), for fields G (top) and C (bottom),
at a common angular resolution. Contours on the H$_3^+$ maps: the S(6) map at 20, 40, 60 and 80\% of its peak; $r$ is the pixel-by-pixel correlation coefficient with the first-order S(6) map from channel 1 (not shown), over the pixels it covers where the H$_3^+$
line is detected at more than $3\,\sigma$\mg{; the bright H$_3^+$ knot at ($+5''$, $-7.9''$) in field C (Section~\ref{sec:detect}) lies outside the channel-1 field and is not included}. \bn{In field G} the color scales extend to slightly negative values so that noise is shown.
Offsets are relative to $\alpha = 22^{\rm h}56^{\rm m}04\fs73$, $\delta = +62\arcdeg02\arcmin15\farcs6$ (G) and
$22^{\rm h}56^{\rm m}02\fs83$, $+62\arcdeg02\arcmin10\farcs5$ (C) (J2000).
\label{fig:maps}}
\end{figure*}

Finally, the H$_3^+$ emission follows the hot shocked H$_2$ (Figure~\ref{fig:maps}). The map of
the $16.325\,\mu$m line closely resembles that of H$_2$ S(6) (\mg{which traces the hot H$_2$ as closely as S(7), used to define the aperture, but whose second-order image in channel 3A covers the full field of the $16.325\,\mu$m map;} correlation coefficient $r = $\,0.97 in G and \bl{0.87} in C, over
pixels where the H$_3^+$ line is detected at more than $3\,\sigma$; all maps \re{used for $r$} are at the common resolution of Section~\ref{sec:obs})\footnote{In Figure~\ref{fig:maps}, S(6) is mapped through its image in second grating order, which a spectral leak admits into MIRI channel 3A
\citep{Neufeld2026}; unlike the first-order line, which lies in channel 1, this image covers the same channel 3 field of view
as the $16.325\,\mu$m line. This leak of second-order light from $\approx 6\,\mu$m into channel 3A is documented by the JWST
pipeline (\url{https://jwst-pipeline.readthedocs.io/en/stable/jwst/spectral_leak/main.html}); the pipeline correction acts only on
1D spectra extracted for compact sources and does not alter the spectral cubes, which therefore retain the S(6) image. All correlation coefficients are computed instead with the first-order S(6) map from channel 1, convolved
to the common resolution, over the part of the field that it covers.}. A map obtained by stacking \re{29} H$_3^+$ lines without \re{coincidences with lines of other species} gives $r = $\,\bl{0.94} with S(6) in G (\bl{0.83} in C) over the pixels where the stack exceeds $3\,\sigma$, but $|r| \le \bl{0.11}$ with
[Fe\,{\sc ii}]\,5.34\,$\mu$m, [Ne\,{\sc ii}]\,12.81\,$\mu$m and H\,{\sc i} 7--6 in G (negative in C).
The correlation with H$_2$ S(1) at 17.03\,$\mu$m, the H$_2$ line closest in wavelength, is weaker ($r = $\,\bl{0.78} in G and \bl{0.68}
in C) than with S(6), as expected if the H$_3^+$ traces the hot H$_2$ rather than the cooler gas that dominates S(1). The H$_3^+$ map
is thus unmistakably that of the hot molecular gas, not of the ionic or atomic emission in the same field.
\mg{The brightest H$_3^+$ emission in field C, a knot at its southern edge ($+5''$, $-7.9''$; Figure~\ref{fig:maps}), is also genuine: in a $0.7''$-radius circle the $16.325\,\mu$m line is detected at \bl{$33\,\sigma$}, with the instrumental width and at the velocity of the second-order S(6) image, and the $(9,0) \to (8,3)$ $11.738\,\mu$m, $(6,3) \to (6,6)$ $17.199\,\mu$m, $(6,1) \to (5,4)$ $12.316\,\mu$m and $(9,1) \to (8,2)$ $13.500\,\mu$m lines are detected at \bn{6--9}\,$\sigma$. Unfortunately the knot lies outside the fields of channels 1 and 2, so that of the H$_2$ lines only S(1), S(2) and the second-order S(6) image cover it. Relative to S(6), the H$_3^+$ line is \bl{2.5} times stronger at the knot than on the ridge to its north-west, while S(6)/S(1) is slightly lower. The knot lies beside a knot of bright, broad (FWHM 140--210\,\kms) and blueshifted ionic emission, in which [Ne\,{\sc ii}], [Ne\,{\sc iii}], [S\,{\sc iii}] and [Fe\,{\sc ii}] are 5--7 times brighter than on the ridge. A plausible explanation is that ionizing radiation (extreme ultraviolet and soft X-rays) from the adjacent fast shock raises the ionization rate of the H$_2$, and hence the production of H$_3^+$.}

\subsection{Excitation} \label{sec:excit}

\begin{figure*}[t]
\plotone{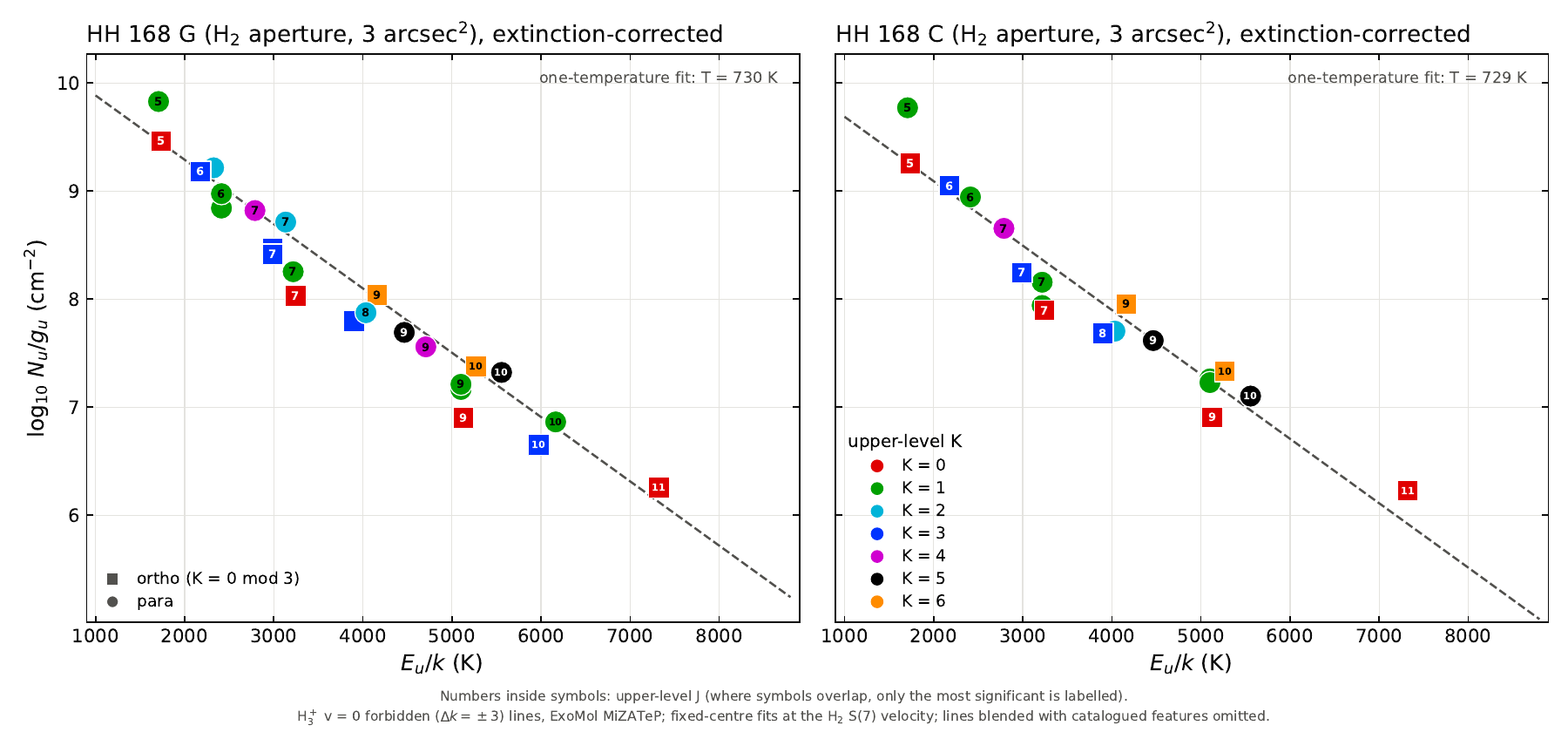}
\caption{Extinction-corrected rotation diagram for the H$_3^+$ lines detected at more than $5\,\sigma$. Colors denote
the upper-level $K$; squares are ortho and circles para levels; the upper-level $J$ is written inside \bn{most symbols}.
Dashed lines: one-temperature fits to the plotted lines (\bn{730}\,K in G, \bn{729}\,K in C).
\label{fig:rotdiag}}
\end{figure*}

Figure~\ref{fig:rotdiag} shows the extinction-corrected rotation diagram. A single\bn{-}temperature fit \mg{to the lines detected at more than $5\,\sigma$} gives
$T_{\rm rot} = $\,\bn{730}\,K (G) and \bn{729}\,K (C), and, extrapolating to unobserved levels in local thermodynamic equilibrium (LTE),
$\log N({\rm H_3^+}) = $\,\bn{$12.70 \pm 0.10$} (G) and \bn{$12.50 \pm 0.11$} (C), corresponding to $N({\rm H_3^+})/N({\rm H_2})
\approx 2 \times 10^{-8}$ and $1 \times 10^{-8}$\re{, relative to the total (warm and hot) H$_2$ column}. However, the \re{LTE} fit is poor ($\chi^2/{\rm dof} = $\,\bn{30 (G) and 35 (C)} with the statistical errors alone, \bn{3.9 (G) and 5.1 (C)} if a 0.1\,dex systematic
uncertainty is added to each line); at a given
energy, levels with large $A$ are systematically underpopulated, as expected if the excitation is subthermal. Subthermal excitation also
explains why $T_{\rm rot}$ is low compared with the collision temperature. When each collisional excitation is followed by radiative decay, $N_u/g_u \propto
e^{-E_u/kT}/A_u$.  The total decay rate of the upper level is found to rise roughly as $\exp(E_u/E_A)$, with
$E_A \approx \bl{1100}$\,K, so that $T_{\rm rot} \approx (1/T + 1/E_A)^{-1}$. The observed $T_{\rm rot}$ thus requires a
collision temperature $T \approx$\,\bn{2200}\,K, and $T_{\rm rot}$ could not exceed \bn{$\approx E_A \approx 1100$}\,K however hot the gas. With critical densities
$A/k_L \sim 10^{6}$--$10^{8}\,\rm cm^{-3}$, subthermal excitation is expected at the densities of shocked gas. Here $k_L = 2 \times 10^{-9}\,\rm cm^3\,s^{-1}$ is the Langevin rate for
H$_3^+$--H$_2$ collisions, $k_L = 2\pi e\,(\alpha/\mu)^{1/2}$ \citep{Gioumousis1958}, with $\alpha$ the polarizability of the
neutral and $\mu$ the reduced mass (for H$_2$, $\alpha = 0.80\,\mbox{\AA}^3$, \citealt{Kolos1967}, giving $1.9 \times 10^{-9}\,\rm cm^3\,s^{-1}$,
which we round to $2 \times 10^{-9}$). Because the ion--induced-dipole attraction makes the capture cross section scale as $1/v$,
$k_L$ is independent of temperature, and it exceeds the product of a geometric cross section and the thermal speed
($\sim 4 \times 10^{-10}\,\rm cm^3\,s^{-1}$ at 2000\,K) by a factor of several.

To interpret the level populations, we solved the equations of statistical equilibrium for the 220 $v = 0$ levels of
H$_3^+$ within $13{,}700\,\rm cm^{-1}$ (19{,}700\,K) of the (1,1) ground level ($J \le 20$; 73 ortho and 147 para), including all 1219
radiative transitions among them from MiZATeP, so that radiative cascades within $v = 0$ are fully included. We neglect vibrational
excitation (the $\nu_2$ fundamental lies at $2521\,\rm cm^{-1}$, or 3630\,K) and any subsequent radiative cascade into the
$v = 0$ levels. At the temperatures inferred below, roughly half of all collisions would leave H$_3^+$ vibrationally excited,
but the ions return to $v = 0$ within $\sim 10$\,ms through the $\nu_2$ band ($A \sim 100\,\rm s^{-1}$), with $|\Delta J| \le 1$ at each step, so that
the rotational distribution of the excitation is largely preserved; the neglect matters mainly for the much hotter collisions
in the shock models (Section~\ref{sec:shocks}). Radiative excitation was also neglected, the infrared radiation field in HH 168
being weak, and the lines were assumed to be optically thin. They are thin by a wide margin: for the $16.325\,\mu$m line in G,
$N_u \approx \bl{1.3} \times 10^{11}\,\rm cm^{-2}$ gives a line-center optical depth \re{$\lesssim 10^{-4}$ (for the lower-level columns of the models below)} for a velocity dispersion of 20\,\kms\ or more,
so that the lines remain thin even along an edge-on path through the shock, and a beam-averaged brightness temperature of
$\approx \bl{50}$\,K, far below the kinetic temperature.

State-to-state rate coefficients for H$_3^+$--H$_2$
collisions are not available for the levels and temperatures of interest; published calculations are limited to low
$J$ and $T \le 300$\,K. Because the H$_3^+$ + H$_2$ reaction proceeds through the deeply bound H$_5^+$ complex, we adopted a
statistical, ``strong-collision'' model: every collision occurs at the Langevin rate $k_L$ (H and He colliders being
weighted by the ratios of their Langevin rates to that for H$_2$, 1.16 and \re{0.42}, for $\alpha = 0.667$ and $0.205\,\mbox{\AA}^3$, \citealt{Schwerdtfeger2019}; these ratios are applied to the adopted $k_L = 2 \times 10^{-9}\,\rm cm^3\,s^{-1}$), and the products are distributed over all levels
in proportion to their thermal populations at the kinetic temperature, $k(i \to f) = k_L\,g_f\,e^{-E_f/kT}/Q(T)$,
independent of the initial level. The name refers to the limit in which every collision forms a long-lived complex that
retains no memory of the initial state, so that a single collision can change the rotational energy by any amount; in the
opposite, weak-collision limit, each collision changes $J$ and $K$ by small steps, with rates that fall off steeply with the
energy transferred (the energy-gap model considered below is of this kind). This model obeys detailed balance and fully scrambles ortho and para H$_3^+$. It fits
the data well ($\chi^2/{\rm dof} = $\,\bl{1.5 (G) and 1.4 (C)}, with the 0.1\,dex systematic
uncertainty; \bn{10 and 8} with the statistical errors alone) for kinetic temperatures of \bn{2300--2750}\,K, \re{higher than that of the hot H$_2$ component in the fit of
Section~\ref{sec:ext} (1300--1400\,K)}.\footnote{\rv{Inelastic collisions with electrons were also considered. Because electrons do not share the ion--neutral drift, they would excite H$_3^+$ at the electron temperature rather than at the drift-enhanced effective temperature of the ion--neutral collisions. Electron-impact rate coefficients have been computed only for levels with $J \le 4$ \citep{Faure2003}; they are of order $10^{-7}\,\rm cm^3\,s^{-1}$ at 1000--3000\,K for $\Delta K = 0$, $|\Delta J| = 1$ and 2 transitions, and about four orders of magnitude smaller for $|\Delta K| = 3$. With rates of this magnitude, electrons compete with neutral collisions at the Langevin rate only if $x_e \gtrsim 10^{-2}$. In the C-type shock model $x_e \lesssim 10^{-5}$ where the H$_3^+$ is excited, so that electron excitation is negligible. In the J-type model, where $x_e \sim 0.1$ in the emitting layer, electron impact (with the rates of \citealt{Faure2003} extrapolated to higher $J$ and $T$) raises the predicted H$_3^+$ emission by $\sim 0.1$\,dex. Because the electrons there share the $\sim 10^5$\,K gas temperature, it slightly flattens the predicted population distribution and does not improve the fit.}}

The strong-collision model is the most efficient possible for large jumps in energy. For H$_3^+$ + H collisions,
close-coupling rate coefficients are available for excitation out of the lowest ortho and para levels at $T \le 300$\,K
\citep{FelixGonzalez2025}. At 300\,K they are within a factor of 2 of the strong-collision values for jumps in energy $\Delta E
\lesssim kT$, but fall below them for larger jumps, approximately as $\exp(-\Delta E/\Delta)$ with $\Delta/k = 459$\,K (rms
deviation 0.14\,dex over 22 transitions). We therefore also considered an energy-gap model, in which each strong-collision
rate coefficient is multiplied by $\exp(-|E_f - E_i|/\Delta)$ (a factor symmetric in the initial and final levels, so that
detailed balance is preserved). With the gap scaled in proportion to the temperature, $\Delta = 1.53\,kT$ (i.e.\ $\Delta/k =
459$\,K at 300\,K), this model fits the HH 168 data equally well, but for higher temperatures (\bn{3250--3750}\,K); with $\Delta$
held at its 300\,K value the observations cannot be reproduced ($\chi^2/{\rm dof} \ge$\,\bl{12}). \mg{Including the lines detected at 3--5\,$\sigma$ changes the temperatures fitted with these models by at most \bn{300}\,K, and including those at 2--5\,$\sigma$ changes $T_{\rm rot}$ by at most \bn{20}\,K.} We conclude that the H$_3^+$ is excited in gas at $T \gtrsim 2000$\,K. In either model most
of the H$_3^+$ resides in unobserved metastable ($J = K$) levels, and the total column density is uncertain by \bn{about three}
orders of magnitude.

A test that does not depend on the collision rates is provided by the radiative lifetimes of the upper levels, $\tau_u = 1/\sum_l
A_{ul}$. \bn{In the subthermal limit introduced above}, $N_u/g_u \propto \tau_u\,e^{-E_u/kT}$; this simple relation
neglects population fed into level $u$ by radiative cascades from higher levels. For the levels
with $\tau_u < 30$\,s, the data agree with this relation for $T = \bn{2700}$\,K (G) and \bn{3300}\,K (C), with $\chi^2/{\rm dof} = $\,\bn{1.2}
and \bl{0.9} when a systematic uncertainty of 0.1\,dex is included. Even the longest-lived levels ($\tau_u \approx \re{300}$\,s) show no
sign of the flattening expected when collisions compete with radiative decay; for collisions at the Langevin rate this implies
$n \lesssim \re{4 \times 10^5}\,\rm cm^{-3}$. The normalization of the relation gives $n\,N({\rm H_3^+}) \approx 10^{19}\,\rm
cm^{-5}$, so that $N({\rm H_3^+}) \gtrsim \re{2} \times 10^{13}\,\rm cm^{-2}$, \re{several times the LTE estimate and tens of times the column in the observed levels}. \re{Several of the
longest-lived levels, the low-$J$ para levels (5,1), (6,1) and (6,2),} lie 0.2--\re{0.6}\,dex above the relation, as expected since they also collect
radiative cascades from the levels above.

\section{Origin of the H$_3^+$ emission: C- or J-type shocks} \label{sec:shocks}

\begin{figure*}[t]
\plotone{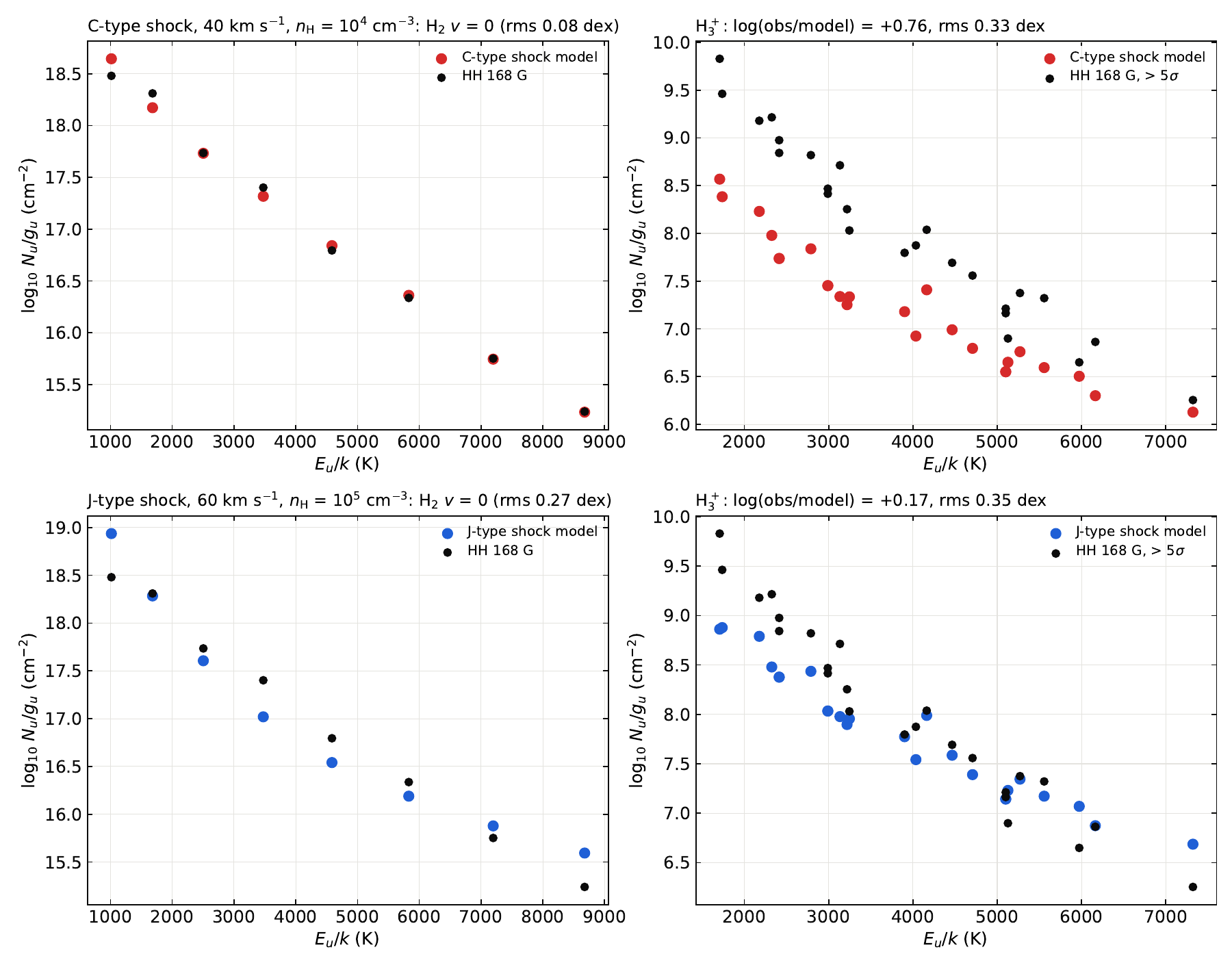}
\caption{Comparison of field G with the best C-type (top, red; \re{40}\,\kms, $n_{\rm H} = 10^4\,\rm cm^{-3}$) and J-type
(bottom, blue; 60\,\kms, $n_{\rm H} = 10^5\,\rm cm^{-3}$) shock models, both for preshock gas irradiated with
$G_0 = 100$ at $A_V = 1$. Black: observations. Left: H$_2$ $v = 0$ rotation diagrams; right: H$_3^+$ lines above
$5\,\sigma$ \rv{(upper-level energies measured from the (1,1) level)}. For each model the area factor $\Phi$ is fitted to the H$_2$ lines; the H$_3^+$ intensities follow with no further
freedom. The H$_3^+$ populations are computed with collision products thermalized at the H$_3^+$--H$_2$ collision
temperature $T_{\rm eff}$. The same comparison for field C is shown in Figure~\ref{fig:compareC}.
\label{fig:compare}}
\end{figure*}

We compared the observations with models computed with the Paris-Durham shock code \citep{Godard2019}, which solves for
the chemistry, including H$_3^+$, along with the H$_2$ level populations. The code does not compute the H$_3^+$ rotational
populations; we computed them at each point in the shock from the code's density, H$_3^+$ abundance and temperatures, with the
strong-collision model of Section~\ref{sec:excit}. The products of each collision were thermalized at the
temperature that characterizes the H$_3^+$--H$_2$ collision energy, $T_{\rm eff} = (m_i T_{\rm n} + m_n T_{\rm i})/(m_i +
m_n) + m_i m_n \Delta v^2/[3k(m_i + m_n)]$, where $T_{\rm n}$ and $T_{\rm i}$ are the neutral and ion temperatures,
$\Delta v$ is the ion--neutral drift speed, and $m_i$ and $m_n$ are the masses of H$_3^+$ and H$_2$. In the J-type shock,
where ions and neutrals are well coupled, $T_{\rm eff} = T_{\rm n}$. For each model, one geometric factor $\Phi$, the ratio of the shock surface area to the projected area of the aperture
\citep{Chernoff1982,Neufeld1989b}, was fitted to the H$_2$ $v = 0$ S(1)--S(8) lines \mg{(Figures~\ref{fig:compare} and \ref{fig:compareC})}; the H$_3^+$ intensities then follow
with no further freedom. The models are computed face-on, so $\Phi = 1$ corresponds to a single plane shock filling the
aperture.

The models start from preshock gas irradiated by a far-ultraviolet field $G_0 = 100$ (Habing units) at $A_V = 1$, with
$B = 3\,(n_{\rm H}/{\rm cm^{-3}})^{1/2}\,\mu$G; the choice of these conditions, the model grid, the H$_3^+$ chemistry and the
sensitivity of the predictions to the uncertain processes are described in Appendix~\ref{app:shocks}, and the structure of the two
best models is shown in Figure~\ref{fig:profiles}.

{\it C-type shock}: The best-fitting C-type shock (Figure~\ref{fig:compare}, top\mg{, and Figure~\ref{fig:compareC}}) has a velocity of \re{40}\,\kms\ and
$n_{\rm H} = 10^4\,\rm cm^{-3}$. It reproduces the H$_2$ $v = 0$ lines with an rms deviation of \re{0.08} (G) and \re{0.11}\,dex
(C), with $\Phi = $\,\re{0.87} and \re{0.77} (here and below, rms values are root-mean-square deviations of $\log({\rm observed/model})$
after the overall scale has been removed, by fitting $\Phi$ for H$_2$ and by subtracting the \re{mean} offset for H$_3^+$\re{, the median offset being}
quoted separately as a factor; they therefore measure disagreements in the relative line strengths only), i.e.\ a single shock surface filling about \re{80\%} of the aperture. The H$_3^+$ forms in warm ($T_{\rm n} \approx$\,\re{1500}\,K), drift-heated molecular gas (Appendix~\ref{app:shocks}).
Because of the ion--neutral drift (up to \re{24}\,\kms), however, the
H$_3^+$--H$_2$ collisions occur at $T_{\rm eff} \approx \re{3 \times 10^4}$\,K, i.e.\ at collision energies of \re{a few} eV,
comparable to the \re{4.3}\,eV needed to dissociate H$_3^+$. With the products thermalized at $T_{\rm eff}$, the model
underpredicts the H$_3^+$ emission by a factor \bn{5.8} (G) and \bn{5.5} (C) and predicts a level distribution flatter than observed
(rms \bn{0.26--0.33}\,dex). The excitation and survival of H$_3^+$ in a C-type shock are therefore
highly uncertain (Appendix~\ref{app:shocks}).

{\it J-type shock}: The best-fitting J-type shock (Figure~\ref{fig:compare}, bottom\mg{, and Figure~\ref{fig:compareC}}) has a velocity of 60\,\kms\ and
$n_{\rm H} = 10^5\,\rm cm^{-3}$. It fits the H$_2$ $v = 0$ lines less well (rms 0.27 and 0.22\,dex), with $\Phi = $\,1.5 and
1.3, and predicts an
H$_3^+$ emission \bn{0.68--0.73} times that observed. Here the H$_3^+$ forms in the hot ($\sim 7 \times 10^4$\,K) gas just behind the
shock front (Figure~\ref{fig:profiles}, right), where H$^+$ + H$_2 \to$ H$_2^+$ + H proceeds rapidly, followed by H$_2^+$ + H$_2 \to$ H$_3^+$ + H, and its
emission shows a flatter dependence on upper-level energy than observed (rms \bl{0.30--0.35}\,dex). \bl{This model lies outside the range for which the shock code has been validated: it includes only singly ionized species, and its documentation restricts J-type shocks to $V_s \le 30$\,\kms. Behind the front the gas reaches $1.8 \times 10^5$\,K and becomes at most 14\% ionized. This ionization is limited by the short time the gas spends hot rather than by the preshock ionization: raising the preshock electron abundance from $4 \times 10^{-6}$ to $10^{-2}$ changes the predicted H$_3^+$ emission by less than 0.01\,dex. Photoionization and photodissociation by the shock's own ultraviolet radiation, which are not included, could however change the abundances of H$^+$ and H$_2$ from which the H$_3^+$ forms, so the J-type predictions are suggestive only. Moreover, \bn{about 80\%} of the predicted H$_3^+$ emission arises in the thin layer ($N_{\rm H} \lesssim 10^{17}\,\rm cm^{-2}$) behind the front in which H$_2$ is still being dissociated, a region not treated in earlier models of fast dissociative shocks \citep{Neufeld1989a}; \bn{beyond $N_{\rm H} \sim 10^{18}\,\rm cm^{-2}$}, where those models apply, both calculations give far lower H$_3^+$ abundances ($\lesssim 10^{-8}$).}

\medskip
The J-type shock can therefore account for the observed H$_3^+$ emission to within a factor $\sim 1.5$, and the C-type shock \re{only} to within a factor \bl{5--6}, without
invoking an enhanced cosmic-ray ionization rate\bl{, although the J-type model extrapolates the shock code beyond its validated range}. A molecular J-type shock at 60\,\kms\ is not extreme in HH 168: the \re{strong [Ne\,{\sc iii}]\,15.56\,$\mu$m emission and the [Ar\,{\sc iii}]\,8.99\,$\mu$m line} show that much faster, ionizing shocks are present just inside the molecular rims of
fields G and C. The C-type shock better reproduces the H$_2$ lines, and the J-type shock
the total H$_3^+$ emission. In both models the H$_3^+$ is excited by collisions at \re{$\sim 3$--$7 \times 10^4$}\,K, and in
both the predicted level distribution is flatter than observed, which suggests that only part of the collision energy is
transferred to H$_3^+$ rotation. Formation pumping and the energy-gap propensity of Section~\ref{sec:excit} change these
predictions by at most \re{about} 0.3\,dex; the remaining discrepancies are comparable to the uncertainties in the excitation by energetic
collisions, in the H$^+$ + H$_2$ charge-transfer rate and in the preshock radiation field, and a mixture of shock conditions, as
expected over an extended shock surface, could well improve the agreement (Appendix~\ref{app:shocks}).

\section{H$_3^+$ in other shocked regions observed with JWST} \label{sec:other}

The forbidden lines should be present wherever hot H$_2$ is bright, but they are weak; even in HH 168 the strongest line
carries only \bl{1/300--1/400} of the flux of H$_2$ S(6). The most promising targets are shocks far from embedded protostars, where the
weak infrared radiation field cannot pump the dense forest of molecular lines (CO$_2$, H$_2$O, C$_2$H$_2$) that crowds the mid-infrared spectra of
irradiated regions and would blend with the H$_3^+$ lines. The bow shocks of HH 211 are an
example: in archival MIRI MRS data with only 200\,s per band \re{at each mosaic pointing} \citep{Caratti2024}, the $16.325\,\mu$m line is detected at
$12\,\sigma$ at the BS3 bow shock and at \bl{$12\,\sigma$} at \mg{BS1,} $6''$ to its east, followed by the \re{$(9,0) \to (8,3)$} \bn{and} \re{$(7,0) \to (6,3)$} \bn{lines}, and the \bl{extinction-corrected} rotation diagrams give $T_{\rm rot} \approx \bl{700}$\,K (Appendix~\ref{app:hh211}). Relative to the hot
H$_2$, the H$_3^+$ emission is 0.3--1 times that of HH 168, although HH 211 is driven by a \bl{low-mass} Class 0 protostar
\citep{Caratti2024} and lacks the high-ionization emission of HH 168 ([Ne\,{\sc ii}]\,12.81\,$\mu$m/H$_2$ S(7) $\approx \bn{0.002}$--0.003, against
\bn{0.10--0.11} in HH 168\re{, both measured at native resolution}, and no [Ne\,{\sc iii}] or [Ar\,{\sc iii}]); this suggests that the H$_3^+$ emission follows the hot H$_2$ largely independently
of the outflow that produces it. \rvb{The pure rotational spectrum of H$_3^+$ is also detected in at least one further protostellar shock, associated with HOPS 78, as we will report separately (S.~Lee, PhD thesis in preparation; S.~Lee et al., ApJ, in preparation).}

\section{Conclusions} \label{sec:conclusions}

\begin{enumerate}
\item We have detected the forbidden rotational spectrum of H$_3^+$ for the first time, toward HH 168 with JWST/MIRI:
\bn{26} lines at more than 5\,$\sigma$ in one or both fields, most prominently the (5,0)\,$\to$\,(4,3) line at $16.325\,\mu$m.
\item The lines share the velocity and spatial distribution of the hot shocked H$_2$. Their level populations are
subthermal, indicating excitation by collisions in gas at $\gtrsim 2000$\,K.
\item Models of C-type and J-type shocks propagating into mildly irradiated molecular gas \re{both reproduce the H$_2$ rotational
lines, the C-type shock more closely}. In both, the H$_3^+$ is excited
by collisions at energies of several eV, for which the excitation is poorly known; this is the main uncertainty in the
predictions, especially for the C-type shock.  Despite this uncertainty, the shock models favor a J-type shock, which reproduces the H$_3^+$ emission to within a factor $\sim 1.5$\bl{, although at 60\,\kms\ this model lies beyond the range for which the shock code is validated}; the C-type shock falls short by a factor of \bl{5--6}, although the excitation and survival of H$_3^+$ in its ion--neutral drift zone are the least certain part of the calculation. \mg{In both models the H$_3^+$ emission comes from the same shock as the hot H$_2$ emission, so the close match of their maps is expected for either type of shock and does not discriminate between them; since neither model fits both species perfectly, the emission may arise from a range of shock conditions over the shock surface.}
\item State-to-state rate coefficients for H$_3^+$--H$_2$ collisions at high $J$ and high collision energy are needed to exploit these lines
as a quantitative probe of the ionization in shocked gas. The lines should be detectable in other outflow and protostellar sources with
strong warm H$_2$ emission and weak infrared irradiation, as HH 211 shows (Section~\ref{sec:other}). \rv{Because H$_2$ and H$_3^+$ are a closely related neutral and ion, observations of both, and in future of heavier ions such as HCO$^+$ at high spectral resolution, offer a means of testing models of the structure of C-type magnetic shocks, in which ions and neutrals move at different velocities.}
\end{enumerate}

\begin{acknowledgments}
This work is based on observations made with the NASA/ESA/CSA James Webb Space Telescope. The data were obtained from the
Mikulski Archive for Space Telescopes at the Space Telescope Science Institute, which is operated by the Association of
Universities for Research in Astronomy, Inc., under NASA contract NAS 5-03127 for JWST. These observations are associated
with program \#7084. Support for program \#7084 was provided by NASA through grant JWST-GO-07084 from the Space Telescope
Science Institute. D.A.N.\ was supported in part by JWST Theory grant JWST-AR-06829, ``Understanding the rich molecular spectra
observed with NIRSpec and MIRI/MRS toward protostellar outflows.'' \rv{We thank the anonymous referee for helpful comments.}

Use of generative AI: Anthropic's Claude (Opus 5.5 and Fable 5.1, accessed through Claude Code) was used for data analysis, coding, running
and analyzing the shock models, generating the figures \mg{from the data using Matplotlib}, \bl{and} \mg{drafting and revising the manuscript text}\bl{.}
OpenAI's Codex was used to write the initial line-fitting and mapping code. The authors take full responsibility for the content of this paper.
\end{acknowledgments}

\facility{JWST (MIRI)}
\software{Paris-Durham shock code, version 1.1 rev.\ 122 \citep[\url{https://ism.obspm.fr};][]{Flower2003,Godard2019}\mg{, Matplotlib \citep{Hunter2007}}}

\startlongtable
\begin{deluxetable*}{lccccc}
\tablecaption{H$_3^+$ forbidden lines detected at more than $5\,\sigma$\label{tab:lines}}
\tablehead{\colhead{Transition} & \colhead{$\lambda$ ($\mu$m)} & \colhead{$A$ (s$^{-1}$)} & \colhead{$E_u/k$ (K)} & \colhead{$F_{\rm G}$} & \colhead{$F_{\rm C}$}}
\startdata
$(10,3) \to (9,6)$ & 7.9497 & $1.1 \times 10^{-1}$ & 5973 & \bl{4.7 $\pm$ 0.9} & \bl{3.3 $\pm$ 1.1} \\
$(8,3) \to (7,6)$ & 8.4099 & $1.6 \times 10^{-2}$ & 3901 & \bl{7.1 $\pm$ 1.0} & \bl{4.9 $\pm$ 0.6} \\
\bl{$(7,3) \to (6,6)$} & \bl{8.7244} & \bl{$3.3 \times 10^{-3}$} & \bl{2990} & \bl{5.5 $\pm$ 1.0} & \bl{4.2 $\pm$ 1.0} \\
$(9,1) \to (8,4)$ & 10.3040 & $7.7 \times 10^{-2}$ & 5101 & \bl{3.5 $\pm$ 0.4} & \bl{3.1 $\pm$ 0.4} \\
$(6,2) \to (5,5)$ & 10.5177 & $1.7 \times 10^{-3}$ & 2325 & \bl{5.4 $\pm$ 0.8} & \bn{3.7 $\pm$ 0.6} \\
$(11,0) \to (10,3)$ & 10.6620 & $5.1 \times 10^{-1}$ & 7322 & \bl{6.2 $\pm$ 0.3} & \bl{5.1 $\pm$ 0.9} \\
$(9,6) \to (9,9)$ & 10.7912 & $2.4 \times 10^{-2}$ & 4163 & \bl{14.8 $\pm$ 0.8} & \bl{10.7 $\pm$ 0.6} \\
\bn{$(7,1) \to (6,4)$\tablenotemark{a}} & \bn{11.4921} & \bn{$1.4 \times 10^{-2}$} & \bn{3218} & \bn{5.3 $\pm$ 0.8} & \bn{3.7 $\pm$ 0.5} \\
$(10,6) \to (10,9)$\bn{\tablenotemark{a}} & 11.4963 & $7.4 \times 10^{-2}$ & 5270 & \bn{10.4 $\pm$ 0.9} & \bn{8.3 $\pm$ 0.7} \\
$(9,0) \to (8,3)$ & 11.7384 & $1.6 \times 10^{-1}$ & 5127 & \bl{6.6 $\pm$ 1.3} & \bl{5.9 $\pm$ 0.8} \\
$(6,1) \to (5,4)$ & 12.3163 & $4.2 \times 10^{-3}$ & 2413 & \bl{7.0 $\pm$ 0.6} & \bl{5.8 $\pm$ 1.0} \\
$(10,1) \to (9,2)$ & 12.6096 & $1.4 \times 10^{-1}$ & 6164 & \bl{2.8 $\pm$ 0.4} & \bl{1.0 $\pm$ 0.3} \\
$(9,5) \to (9,8)$ & 12.9717 & $4.2 \times 10^{-2}$ & 4465 & \bl{5.2 $\pm$ 0.9} & \bl{4.0 $\pm$ 0.7} \\
$(5,1) \to (4,4)$ & 13.3642 & $7.5 \times 10^{-4}$ & 1707 & \bl{7.3 $\pm$ 0.6} & \bl{5.8 $\pm$ 0.7} \\
$(7,0) \to (6,3)$ & 13.4582 & $3.2 \times 10^{-2}$ & 3246 & \bn{13.6 $\pm$ 1.0} & \bn{9.1 $\pm$ 0.6} \\
$(9,1) \to (8,2)$ & 13.4998 & $7.2 \times 10^{-2}$ & 5101 & \bl{2.6 $\pm$ 0.4} & \bl{3.0 $\pm$ 0.4} \\
$(10,5) \to (10,8)$ & 13.7123 & $8.3 \times 10^{-2}$ & 5557 & \bn{4.8 $\pm$ 0.4} & \bn{2.7 $\pm$ 0.3} \\
$(7,4) \to (7,7)$ & 14.2794 & $6.8 \times 10^{-3}$ & 2789 & \bn{8.5 $\pm$ 0.6} & \bn{5.4 $\pm$ 0.3} \\
\bl{$(9,4) \to (9,7)$} & \bl{15.7859} & \bl{$4.4 \times 10^{-2}$} & \bl{4707} & \bn{3.5 $\pm$ 0.4} & \bn{2.2 $\pm$ 0.4} \\
\bl{$(7,1) \to (6,2)$} & \bl{16.1011} & \bl{$1.5 \times 10^{-2}$} & \bl{3218} & \bl{2.7 $\pm$ 0.7} & \bl{2.0 $\pm$ 0.3} \\
$(5,0) \to (4,3)$ & 16.3254 & $3.0 \times 10^{-3}$ & 1737 & \bl{21.2 $\pm$ 0.3} & \bl{12.1 $\pm$ 0.2} \\
$(6,3) \to (6,6)$ & 17.1985 & $3.0 \times 10^{-3}$ & 2177 & \bl{12.2 $\pm$ 0.7} & \bl{8.3 $\pm$ 0.4} \\
$(8,2) \to (7,1)$ & 17.6053 & $2.6 \times 10^{-2}$ & 4035 & \bl{3.3 $\pm$ 0.3} & \bl{2.1 $\pm$ 0.4} \\
$(7,3) \to (7,6)$ & 18.0018 & $9.9 \times 10^{-3}$ & 2990 & \bl{7.7 $\pm$ 1.2} & \bl{4.8 $\pm$ 0.9} \\
\bl{$(6,1) \to (5,2)$} & \bl{18.0569} & \bl{$5.3 \times 10^{-3}$} & \bl{2413} & \bl{4.8 $\pm$ 0.8} & \bl{2.8 $\pm$ 0.7} \\
\bl{$(7,2) \to (6,1)$} & \bl{19.9567} & \bl{$1.0 \times 10^{-2}$} & \bl{3134} & \bl{7.4 $\pm$ 1.3} & \bl{4.7 $\pm$ 1.6} \\
\enddata
\tablenotetext{a}{\bn{Fitted jointly; the two lines are 110\,\kms\ apart.}}
\tablecomments{Rest wavelengths and $A$ values from MiZATeP \citep{Mizus2017} with MARVEL energies \rvb{\citep{Bowesman2023}}\rv{; $E_u$ is measured from the (1,1) level}. $F_{\rm G}$, $F_{\rm C}$:
observed (not extinction-corrected) line fluxes in fields G and C, integrated over the $\approx 3\,\rm arcsec^2$ apertures of
Section~\ref{sec:obs}, in units of $10^{-17}\,\rm erg\,cm^{-2}\,s^{-1}$,
from fits of fixed center and width (\bl{the width measured in each sub-band}; statistical uncertainties\bn{; neighboring lines within the fit window fitted as additional components}), summed over the aperture in maps at a common angular resolution (Section~\ref{sec:obs}). Lines blended with other species are omitted, including $(5,2) \to (5,5)$ at $21.830\,\mu$m, which coincides with [Ar\,{\sc iii}]\,$21.829\,\mu$m. $K$ is the
absolute value of the signed quantum number $k$; the selection rule $\Delta k = \pm 3$ allows $|\Delta K| = 1$ when $k$ changes
sign, e.g., $(8,2) \to (7,1)$ is $k = +2 \to -1$.}
\end{deluxetable*}

\clearpage
\appendix

\section{Field C} \label{app:lines}

\begin{figure*}[h!]
\plotone{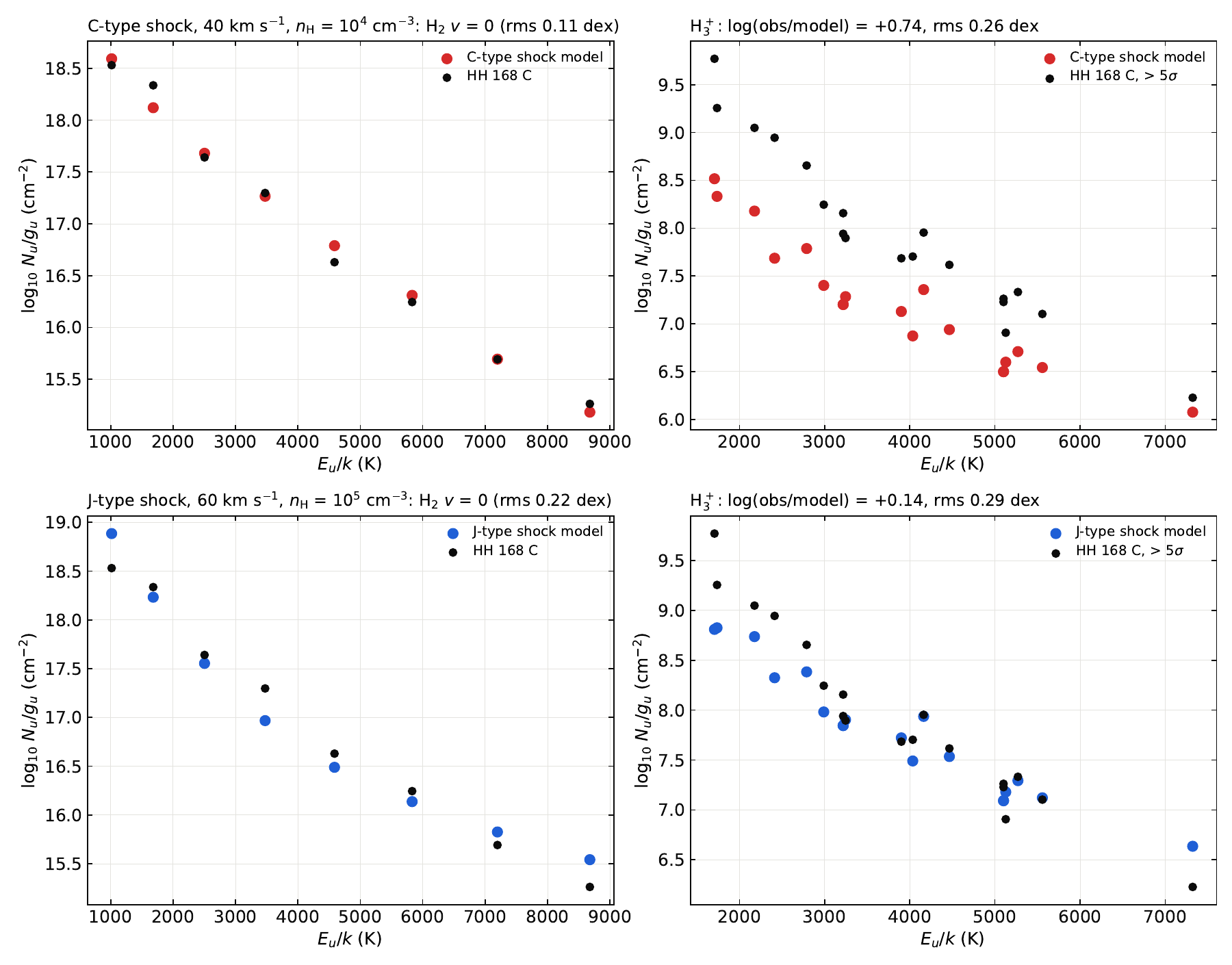}
\caption{As Figure~\ref{fig:compare}, for field C.
\label{fig:compareC}}
\end{figure*}

Figure~\ref{fig:compareC} shows the comparison of field C with the best C- and J-type shock models of
Section~\ref{sec:shocks}.

\clearpage
\section{Details of the shock models} \label{app:shocks}

{\it Preshock conditions and model grid.} The preshock ionization proved to be important. In dark, well-shielded preshock gas the ionization fraction is
$\sim 10^{-7}$, and C-type shocks then underpredict the H$_3^+$ emission by more than an order of magnitude. We therefore computed
the preshock composition self-consistently, as the steady state of gas exposed to a far-ultraviolet field $G_0$
(in Habing units) at visual extinction $A_V$, and then computed the shock with that composition. \re{The best fits to the H$_2$ $v = 0$ lines are obtained from a grid of nine models ($n_{\rm H} = 10^4$, $3 \times 10^4$ and
$10^5\,\rm cm^{-3}$; 40, 50 and 60\,\kms) starting from preshock gas with $G_0 = 100$ and $A_V = 1$} (a field 100 times the \bn{Habing estimate of the} mean interstellar
radiation field, modest compared with the $10^3$--$10^5$ of bright photodissociation regions), in which the electron abundance is
$x_e = 2 \times 10^{-5}$ for $n_{\rm H} = 10^4\,\rm cm^{-3}$ and $3.6 \times 10^{-6}$ for $n_{\rm H} = 10^5\,\rm cm^{-3}$.
\re{Other preshock conditions were also explored: seven 40\,\kms\ models with $G_0 = 10$--$10^4$ and $A_V = 0.1$--1, ten from an
imposed, fully molecular composition with $x({\rm C^+}) = 8 \times 10^{-5}$, and two from dark-cloud gas; none fits the H$_2$ lines
better than the best models of the grid.} \re{The best C-type model, at 40\,\kms, lies at the low-speed edge of this grid; slower shocks are discussed below.}

The magnetic field is $B = b\,(n_{\rm H}/{\rm cm^{-3}})^{1/2}\,\mu$G with $b = 3$ in all models, rather than the commonly
adopted $b = 1$; this value is an assumption, not derived from the observations, and is motivated as follows. For the preshock \re{conditions} considered here, a shock becomes J-type above a critical speed of
$\approx$\,19\,$b$\,\kms, the \re{magnetosonic speed of the charged fluid} \bl{\citep{Draine1980}}\re{, whose inertia is dominated by the charged grains} \bl{\citep{Flower2003}}. With $b = 1$, any shock fast enough to produce the hot H$_2$ would therefore be J-type; a C-type
shock at \re{40}\,\kms, with the large ion--neutral drift that drives the H$_3^+$ chemistry, requires $b \gtrsim \re{2}$\re{, and $b = 3$ keeps shocks
up to $\approx 58$\,\kms\ C-type}. Even with
$b = 3$, the 60\,\kms\ shock exceeds the critical speed and is J-type. We did not explore larger values of $b$: a stronger
field would broaden the C-type shock and lower its peak temperature, so that still faster shocks would be needed to produce the
hot H$_2$. Without searching model space further, we report the most physically plausible parameters. The ultraviolet field enters only
through the preshock composition; within the shock itself it was switched off, so photoionization and photodissociation in
the shocked gas are neglected.

\begin{figure*}
\plotone{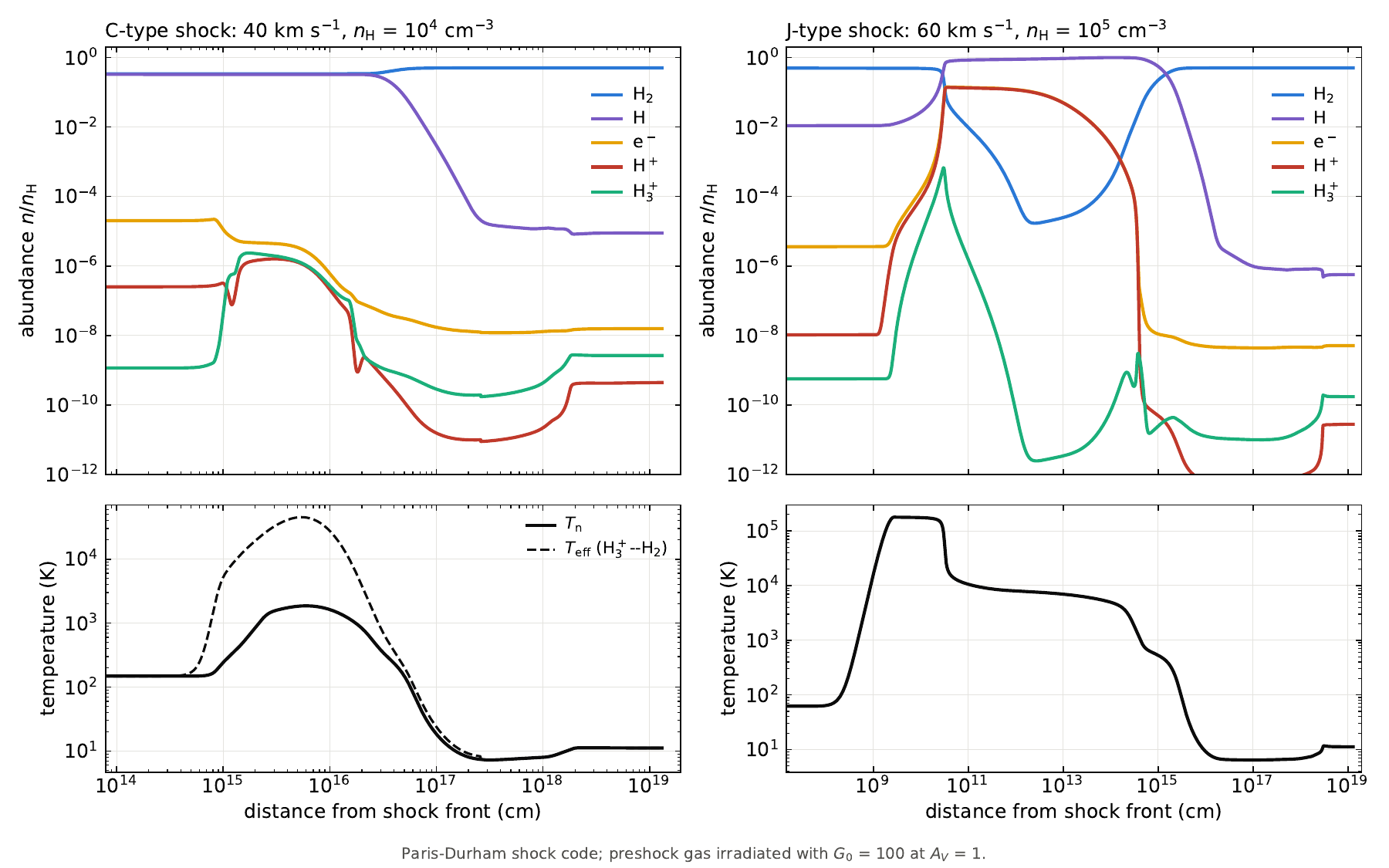}
\caption{Structure of the best C-type (left) and J-type (right) shock models: abundances relative to H nuclei (top) and
temperature (bottom) against distance from the shock front: neutral temperature $T_{\rm n}$ (solid) and, for the C-type
shock, the H$_3^+$--H$_2$ collision temperature $T_{\rm eff}$ (dashed).
\label{fig:profiles}}
\end{figure*}

{\it H$_3^+$ chemistry and excitation.} In the best C-type shock, the H$_3^+$ forms in warm ($T_{\rm n} \approx \re{1500}$\,K), drift-heated molecular gas (Figure~\ref{fig:profiles},
left) by H$_2^+$ + H$_2$, in rapid exchange with H$^+$ and H$_2^+$, which together carry most of the charge there; its
abundance therefore follows the preshock ionization. The charge accumulates in H$_3^+$ because, in hot gas, the
proton-transfer reactions that normally destroy it, H$_3^+$ + X $\to$ XH$^+$ + H$_2$ (X = CO, O, H$_2$O, CH$_4$, \ldots),
are driven in reverse once the collision energy exceeds their endothermicities of $\sim$0.5--3\,eV (cf.\ the analogous reaction of vibrationally excited H$_2$ with HOC$^+$; \citealt{Goicoechea2025}):
since H$_2$ is far more abundant than any X, the positive charge then piles up in H$_3^+$. In the C-type shock the
required energy is supplied by the ion--neutral drift, and in the J-type shock by the high temperature behind the front.

The C-type predictions of Section~\ref{sec:shocks} thermalize the collision products at $T_{\rm eff} \approx$ \re{$3 \times 10^4$}\,K. Thermalizing the products at $T_{\rm n}$ instead would give \bn{an underprediction by a factor of 7.0 and 5.9} and an rms of
\bn{0.18--0.21}\,dex, but has no physical basis when the ions drift through the neutrals. In reality, the energy of each
collision is shared between translation and the internal modes of both products, and collisions at these energies may
also excite H$_3^+$ vibrationally or dissociate it; none of these processes has been calculated. In the chemical network,
collisional dissociation of H$_3^+$ is included only through the H$_2^+$ + H channel (6.2\,eV), with an estimated rate; the
lower-energy H$^+$ + H$_2$ channel (\re{4.3}\,eV) is absent.

Slower C-type shocks (20--30\,\kms), in which the drift is
smaller, predict steeper H$_3^+$ level distributions (rms \bl{0.13--0.32}\,dex), but they \re{reproduce the H$_2$ lines poorly} (rms
\re{0.3--1.6}\,dex for the $v = 0$ lines) and would require $\Phi \approx$\,\bl{3--45} to supply the
observed H$_3^+$ emission, compared with $\Phi \approx$\,\re{5} for the \re{40}\,\kms\ model \re{(whose H$_2$ lines require $\Phi \approx 0.8$). These runs,
with $b = 3$, start from an imposed, fully molecular preshock composition with $x_e = 8 \times 10^{-5}$ rather than the $G_0 = 100$
steady state}; the density has little effect on the
H$_3^+$ distribution, as expected for subthermal excitation.

{\it Sensitivity of the predictions.} With the energy-gap model of Section~\ref{sec:excit} ($\Delta = 1.53\,kT$), the
predictions change little at these temperatures (by \re{0.05} and \re{0.08--0.09}\,dex for the C- and J-type shocks). Because H$_3^+$ is
destroyed and re-formed rapidly in both models (\re{in the C-type shock} mainly through H$_3^+$ + H $\to$ H$_2^+$ + H$_2$ followed by H$_2^+$ + H$_2$
$\to$ H$_3^+$ + H\re{; in the J-type shock it is destroyed \bl{about half by dissociative recombination and half by H$_3^+$ + H $\to$ H$_2^+$ + H$_2$ and electron-impact dissociation,} and re-formed from the H$_2^+$
produced by H$^+$ + H$_2$}), formation pumping is also significant, with a formation rate per ion of \re{0.15--0.2} times the Langevin
collision rate in the C-type shock and 0.7--0.9 times in the J-type shock. Including it, with the nascent H$_3^+$ distributed
thermally at 3000--$10^4$\,K, raises the predicted H$_3^+$ emission by \re{0.07--0.10}\,dex (C-type) and \re{0.26--0.31}\,dex (J-type),
but leaves the predicted level distributions flatter than observed. The remaining
discrepancies are comparable to the uncertainties in the H$_3^+$ excitation by energetic collisions, in the H$^+$ +
H$_2$ charge-transfer rate for the excited H$_2$ present in the shocked gas \re{(state-resolved rates have been calculated for $v = 0$, $J = 0$, \citealt{Savin2004}, and for
vibrationally excited H$_2$, \citealt{Goicoechea2025}, but not for the rotationally hot H$_2$ of a shock)}, and in the preshock radiation field. They should also
be judged against the simplicity of the comparison: each model is a single planar shock of one velocity and density, whereas
the aperture certainly contains a range of shock conditions. Since slower shocks predict steeper H$_3^+$ level distributions,
a mixture of shocks could well improve the agreement, and is a natural assumption for an extended, overlapping shock surface
such as that of HH 168; we defer such models to a later paper.

\section{HH 211} \label{app:hh211}

\begin{figure*}
\plotone{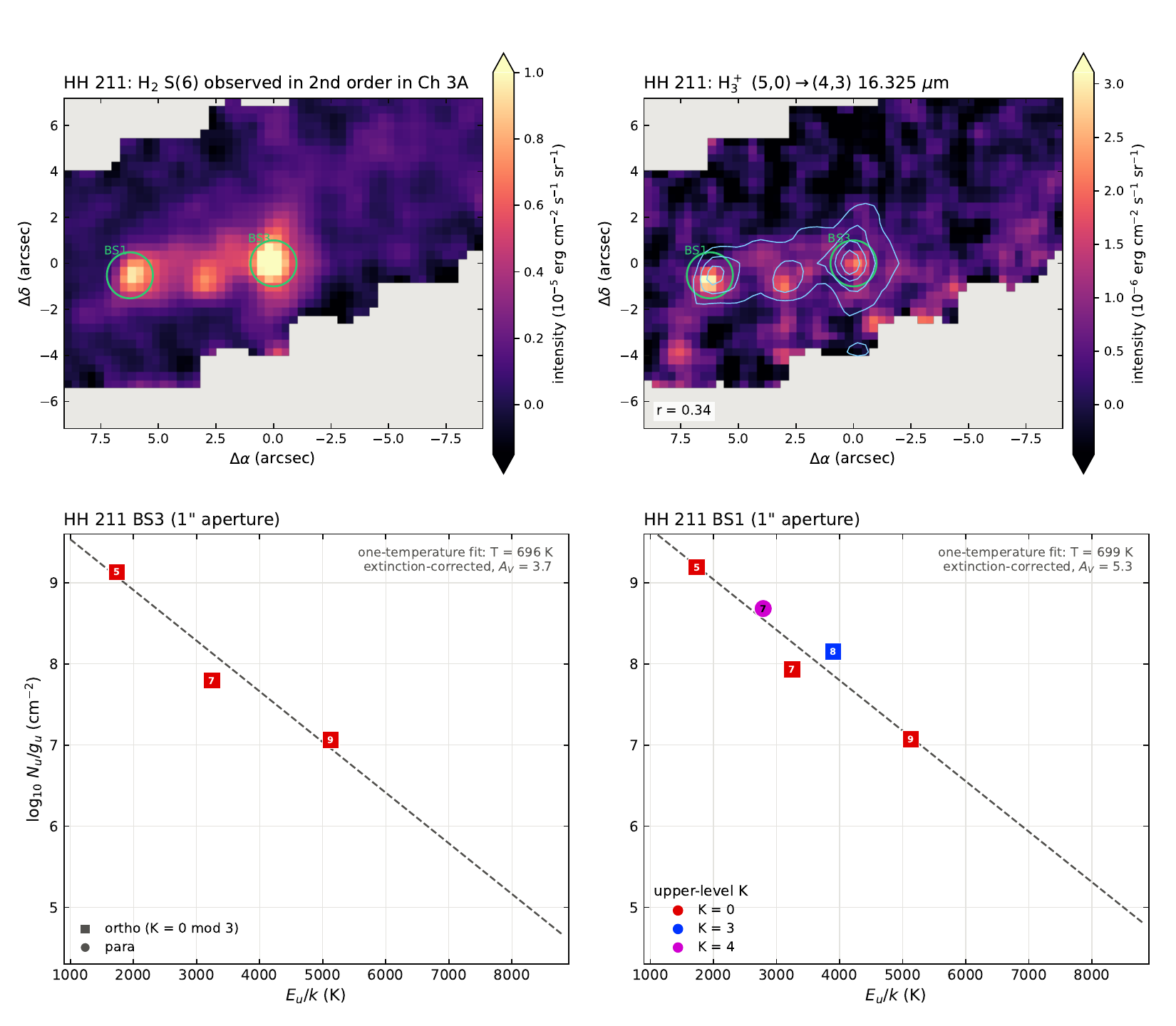}
\caption{H$_3^+$ in HH 211. Top: H$_2$ S(6), observed in second order in channel 3A as in Figure~\ref{fig:maps} (left) and H$_3^+$ $(5,0) \to (4,3)$ $16.325\,\mu$m (right) around the BS3
bow shock \bn{(intensities integrated over $\pm 150$\,\kms\ of the line center; offsets relative to BS3\rv{, at $\alpha = 03^{\rm h}43^{\rm m}59\fs41$, $\delta = +32\arcdeg00\arcmin35\farcs2$ (J2000)})}, matched to a common point-spread function of $0.87''$ and regridded to $0.35''$; contours show S(6) at 20, 40, 60 and 80\% of its peak, circles the two
apertures, and $r$ the correlation of the two maps over pixels where the H$_3^+$ line exceeds $3\,\sigma$. Bottom: \bl{extinction-corrected} rotation
diagrams of the lines detected at more than $5\,\sigma$ in each aperture, with one-temperature fits; symbols as in
Figure~\ref{fig:rotdiag}.
\label{fig:hh211}}
\end{figure*}

Figure~\ref{fig:hh211} shows the H$_3^+$ emission in the archival MIRI MRS data of HH 211 (program 1257; stage-3 cubes from
MAST), analyzed \re{with the line list and fitting method of Sections~\bn{\ref{sec:obs} and \ref{sec:detect}}, but on the native-resolution cubes}: $1''$-radius apertures at the BS3 position of
\citet{Caratti2024} and at \mg{BS1 of the same authors,} $6''$ to its east, fixed-center fits at the H$_2$ velocity\bl{, and an extinction correction from a two-temperature fit to the H$_2$ S(1)--S(8) lines of each aperture as in Section~\ref{sec:ext} ($A_V = 3.7$ at BS3 and 5.3 at BS1). The H$_2$ lines of HH 211 have the instrumental width, but the $16.325\,\mu$m line is broader ($115 \pm 8$\,\kms\ FWHM, against 95\,\kms\ for H$_2$ S(1)); the H$_3^+$ lines were therefore fitted with an intrinsic width of 65\,\kms\ added in quadrature to the instrumental width.}

\bibliographystyle{aasjournal}
\bibliography{refs}

\end{document}